\documentclass[a4paper,twocolumn,10pt]{article}

\newcommand{\ket}[1]{| \, #1 \rangle}

\newcommand{\bra}[1]{\langle #1 \, |}

\def\>{\rangle}
\def\<{\langle}

\newcommand{\be}{\begin{equation}}
\newcommand{\ee}{\end{equation}}
\newcommand{\bea}{\begin{eqnarray}}
\newcommand{\eea}{\end{eqnarray}}

\usepackage{graphicx,subfigure}
\usepackage{amsmath}
\usepackage{amsfonts,amssymb,graphicx,latexsym,color}
\usepackage{bm,bbm}

\title{\flushleft\bf \textsf{Feedback control in quantum optics: an overview of experimental breakthroughs 
and areas of application}}
\author{\hspace*{-15.7cm}\normalsize Alessio Serafini\footnote{{\sf serale@theory.phys.ucl.ac.uk}}\smallskip\\
\hspace*{-8cm}\normalsize Department of Physics \& Astronomy, University College London,\\ 
\hspace*{-9.95cm}\normalsize Gower Street, London WC1E 6BT, United Kingdom}
\date{\hspace*{-15.05cm}\normalsize November 22, 2012\bigskip\\
\hspace*{-3.5cm}\begin{minipage}[h!]{14cm} 
We present a broad summary of research involving the application of quantum feedback control techniques  
to optical set-ups, from the early enhancement of optical amplitude squeezing to 
the recent stabilisation of photon number states in a microwave cavity, 
dwelling mostly on the latest experimental advances.
Feedback control of quantum optical continuous variables, quantum non-demolition memories, 
feedback cooling, quantum state control, adaptive quantum measurements and coherent feedback strategies will all be touched upon in our discussion.
\end{minipage}}

\begin{document}

\maketitle

\noindent Quantum control is a broad field of study, engaging the engineering, mathematics, and physical sciences communities in an effort to analyse, design and experimentally demonstrate techniques whereby the dynamics of physical systems operating at the quantum regime is steered towards desired aims by 
external, time-dependent manipulation \cite{dalessandro08,wiseman10}. The development of advanced quantum control schemes is clearly central to the areas of quantum and nano-technologies, whenever the main focus is on the exploitation of coherent quantum effects. Prominent among classical and quantum control techniques are the so-called feedback or closed-loop 
techniques \cite{wiseman10,zhou96}, where the manipulation applied on the system at a given time depends on its state in the past. Closed loop quantum control is particularly well suited to fight decoherence (the nemesis of quantum information processing, whereby 
the system quantum coherence is lost through unwanted interaction with a large 
macroscopic environment) and stabilise quantum resources in the face of noise. It is therefore a very promising paradigm, which is attracting considerable attention. 

Due to the high degree of coherent control, to the wide availability of well established experimental techniques, and to the relatively low technical noise and decoherence enjoyed by optical set-ups,
the quantum optics community has been in a position to pioneer most of the quantum control techniques developed so far, 
and is still definitely at the forefront of such research.
In particular, quantum optics allows for fast and relatively efficient detections in the quantum regime, for manipulations by control fields on time-scales much shorter than the system's typical dynamical time-scales, as well as for efficient input-output interfaces (as for travelling modes impinging on optical cavities).
These advantages make quantum optical systems particularly well suited for the implementation of closed-loop (`feedback') 
control techniques, where some (classical or quantum) information is extracted from the system and used to condition the 
control operations. Feedback control is, in a sense, the next step in quantum control techniques, with the 
remarkable possibility of stabilising specific target states in the face of decoherence and noise. Also, feedback 
techniques can be applied to cool a number of diverse quantum degrees of freedom, and thus to help bringing them into the quantum 
regime. Over the last decade, several quantum optical demonstrations of feedback techniques have been 
achieved, climaxing with the recent, spectacular stabilisation of photon number states in a cavity QED set-up \cite{sayrin11}, 
and enhanced optical phase estimation \cite{yonezawa12}. 

Our overview will follow a combined historical-contextual order. We shall start with the introduction of some basic terminology 
and notions concerning quantum feedback; next, we will move on to consider the measurement-based feedback control 
of quantum continuous variables in general, and then specialise our treatment to the important cases of quantum memories 
for continuous variables and feedback cooling; we will hence hit the deep quantum regime by considering the 
feedback control of highly coherent systems with few quantum excitations in cavity QED and linear optics set-ups;
we will then review advances in adaptive measurement techniques in the context of optical quantum metrology;
our last stop will be to consider the notion and experimental achievements in the subarea of coherent feedback;
finally we shall include some cursory outlook on research in the area. 
Note that we will emphasise primarily experimental achievements with quantum optical systems and, although partial 
reference to the accompanying theoretical literature will be provided, 
no attempt will be made at a comprehensive coverage of the vast general 
theoretical literature on quantum feedback control.
For that, the reader may refer to textbooks with a broader scope \cite{dalessandro08,wiseman10}. 
Also, we shall not cover the area of molecular control by pulse-shaping driven by 
adaptive feedback and evolutionary algorithms, 
for which the reader is referred to \cite{rabitz09}.

\section{\Large \bf \textsf{Basic concepts and terminology}}

We introduce here the fundamental notions required for a basic understanding of feedback control in quantum systems.
We will limit ourselves to defining and briefly sketching some terminology and typical issues encountered. 
A  detailed, systematic treatment, may be found in \cite{wiseman10,gardiner04,charmichael07}. 
A good introduction to continuously monitored quantum systems, including specific applications in quantum optics 
(fluorescence monitoring), is given in \cite{jacobs06}. See also \cite{toth12} for a very recent overview of the theory.

Ideally, a closed quantum system can be prepared in a pure state vector $\ket{\psi(0)}$ 
(adopting Dirac notation), and then evolves unitarily under the action of some hermitian Hamiltonian 
operator $\hat{H}$, according to $\ket{\psi(t)} = \exp(-i\hat{H}t)\ket{\psi(0)}$ ($\hbar=1$ throughout the article). 
In practice, except for a few exceptions in very special situations 
(typically polarisation degrees of freedom over very short transmission links), quantum degrees of freedom in the 
lab are not described by pure states, but rather by statistical mixtures of pure states, represented as 
$\sum_{j} p_j \ket{\psi_j}\bra{\psi_j}$. Even if the initial state is pure, the microscopic quantum system will interact with its bulky 
environment where part of the quantum information will, in most cases, irreversibly leak, thus ending up in a mixed state 
(a statistical mixture). However, there are situations where the environment of a quantum system can be accessed and 
monitored, in an attempt to retrieve the quantum information leaking out of the system 
(just consider, for instance, the monitoring of light leaking out of a cavity to obtain information about the state inside the cavity). 
In some cases, as we will see, the 
environment interacting with the system can even be engineered 
and used as a `meter', through which the quantum system can be indirectly monitored 
(thanks to the correlations that the environment built up with the system during their previous interaction). 

Assuming, for simplicity, that the global system-environment state $\ket{\psi(0),\psi_E}=\ket{\psi(0)}\otimes\ket{\psi_E}_{E}$ 
was separable and pure at the beginning of the evolution, one has, after a time $t$, $\ket{\psi(t)}=\exp(-i\hat{H}_I t)\ket{\psi(0),\psi_E}$, 
where $\hat{H}_I$ is the interaction Hamiltonian between system and environment. 
Now, by projectively measuring the environment in the basis $\ket{j}_{E}$, one conditionally project (upon the occurrence of outcome $j$) the system on the (non-normalised) state $\bra{j}_E\exp(-i\hat{H}_I t)\ket{\psi(0),\psi_E}$. 
In other words, the state of the system is kept pure and, up to normalisation, 
is acted upon by the linear operator $M_j=\bra{j}_E\exp(-i\hat{H}_I t)\ket{\psi_E}_E$:
$\ket{\psi}\rightarrow M_j \ket{\psi}/\sqrt{\bra{\psi}M_j^\dag M_j\ket{\psi}}$, with 
probability $\bra{\psi}M_j^\dag M_j\ket{\psi}$
This simple model actually describes the most general possible quantum measurement, also known as `positive operator valued measure' (POVM) \cite{peres93}, 
where the environment acts as a correlated meter. If the measurement outcome is not recorded 
(or if such recording is not possible in practice, as is the case with most noisy environments), the system's state evolves to 
a mixed statistical average encompassing all possible measurement results: 
$\ket{\psi}\rightarrow \sum_{j} M_j\ket{\psi}\bra{\psi}M_j^\dag $. Constantly measuring ({\em i.e.}, ``monitoring'') the environment thus prevents the mixing due to the system-environment interaction and allows one to `unravel' the 
open system dynamics into a set of pure state `quantum trajectories'. The unravelling, 
always related to a measurement process, and the ensuing quantum trajectories are key concepts in the theory of 
monitored open quantum systems. Notice that, while the {\em conditioned} state (after a measurement) does depend on the 
particular measurement chosen, the average {\em unconditioned} state must obviously be independent from such a choice, 
and represents the deterministic open system dynamics. 

The conditional evolution, being dependent on the outcome of quantum measurements, involves a fundamental probabilistic element. 
In order to describe such a dynamics, one has hence 
to define appropriate stochastic increments (if the monitoring is continuous) or `quantum jumps' \cite{plenio98,breuer04}. 
The evolution of the conditioned, pure states is then governed by a stochastic Schr\"odinger equation, involving a deterministic part and stochastic increments (governed by Ito calculus, see \cite{gardiner04} for details). When averaged over all the possible realisations of the 
stochastic elements, the stochastic Schr\"odinger equation reduces to a deterministic master equation describing the evolution of the mixed state of the system.
The stochastic Schr\"odinger equations which differ only by the choice of measurement carried out on the environment 
(in the system-meter picture sketched above) 
reduce to the same master equation, solely determined by the form of the interaction between system and environment, and are said 
to provide different quantum unravellings of such a master equation. 

This probabilistic element in the dynamics, whereby the quantum trajectories ({\em i.e.}, the pure conditional states)
evolve continuously but not differentiably in time, 
is the expression of the fundamental probabilistic nature of quantum states, and is referred to as the `back-action' noise induced by the measurement process. 
In this context, especially in relation to quantum control, it is relevant to introduce the notion of `strength' of a quantum measurement. 
The strongest possible class of measurements is represented by projective measurements, where the operators $M_j$ defined above are projectors on a basis of orthogonal states of the system $\{\ket{e_j}\}$. Such measurements provide one with the maximum amount of information on the state, but are also maximally destructive in the sense that,
once the measurement occurs, the system abruptly jumps to the new pure state $\ket{e_j}$ (von Neumann postulate), 
often erasing the system's state before the measurement (which merely influences the probability of outcome $j$).
The weakest conceivable measurement is instead the one where all the $M_j$ are multiple of the identity operator. 
Such a (trivial!) measurement process corresponds to leaving the quantum system alone, and does not reveal any information 
about the system (in fact, if the system state is unknown, it will still be completely unknown after the measurement), but is also 
clearly the measurement implying minimal disturbance. Non-projective POVMs interpolate between these two extreme classes 
of measurements. In general, for a POVM, the greater the information gained about the state of the system, the greater the 
disturbance induced, a heuristic principle which is customarily referred to as the 
`information-disturbance trade-off' \cite{banaszek01}. 
For the purpose of realising measurement-based feedback control, with the typical aim of stabilising quantum states in the face 
of noise and decoherence, repeated `weak' measurements \cite{aharonov90}, where the information gain is relatively low but the state remains largely undisturbed, are typically a suitable choice.\footnote{Although it must be noted that there are 
strategies, based on the so called Zeno dynamics \cite{misra77}, where repeated (strong) projective measurements have been proposed as a way to keep a state in a target pure state, without the need of any other active manipulation \cite{raimond12}.} 
In the system-meter model which was briefly reviewed above, where the meter and the system interact and the meter is then projectively measured, the strength of the measurement is strictly related to the strength of the correlations between system and meter (the `environment' in the argument above). If the system and the meter are in a separable, product state, then measuring the meter does not provide one with any information whatsoever about the system, and one is reduced to the trivial weakest case. If the system and the environment are instead maximally correlated (`maximally entangled'), then a projection of the environment on a properly chosen basis will realise a non-destructive projective measurement on the system as well 
(a so called `non-demolition' measurement \cite{braginsky80}). 
Other intermediate cases (weak measurements and POVMs) can be formally obtained for other choices of the basis and other entangled states. 
In practice, let us notice that sometimes the feasible measurements might not be represented by rank 1 projection to start with 
(as would be the case for an avalanche photodiode which just distinguishes between zero photons and any positive number of photons), and these arguments must then be carefully revisited in such cases. 
They do, however, provide valid general guidelines to understand the central issues involved 
in the study of monitored quantum systems.

After the monitoring, measurement-based feedback loops are typically closed by actuators that apply coherent 
quantum manipulations on the system where some parameter depends on the measurement outcome at the previous step.
Ideally, such actuators act on time-scales which are fast with respect to the system dynamics (and so do the detectors).  

\begin{figure}[t!]
\begin{center}
\subfigure[Measurement-based feedback loop]{{\includegraphics[scale=1]{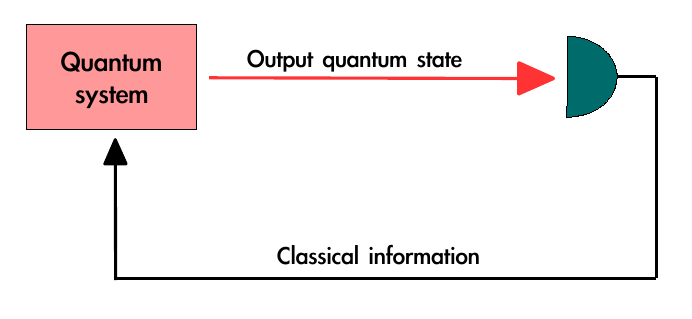}}}\medskip

\subfigure[Coherent feedback loop]{\hspace*{.6cm}{\includegraphics[scale=1]{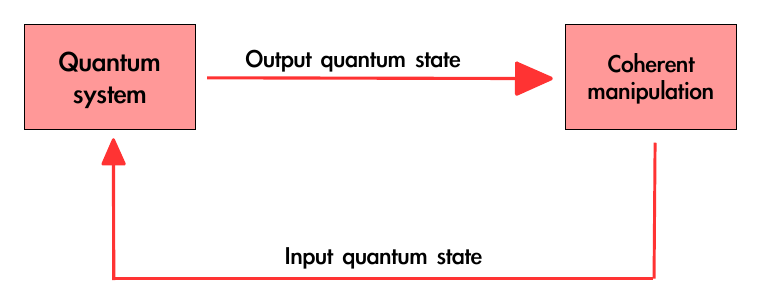}}}
\caption{\small Depiction of measurement-based (a) and coherent (b) feedback loops. In diagram (a) 
the output system, after having interacted with the system, is measured, and the classical information contained in 
the measurement outcome is then used to affect the evolution of the quantum system (typically by modifying the Hamiltonian); 
in diagram (b), the output quantum state is coherently manipulated (typically by a unitary operation), and then 
fed back into the quantum system as an input state through a coherent coupling. \label{types}}
\end{center}
\end{figure}

In this article, we will refer to {\em state} control, whenever the feedback loop is aimed at obtaining, and possibly stabilizing, 
a particular target state, or to optimise a given resource (like quantum correlations or purity). This is opposed to {\em operator} 
control, where the objective of the control procedure is to reproduce a given quantum operation 
on the system (irrespective of the initial state).
We shall also distinguish between `measurement-based' feedback, 
where the system is monitored and then manipulated with operations that depend on the measurement 
outcome (such that the information gleaned from the system and used in its subsequent manipulation is classical) 
and `coherent' feedback, where the system interacts coherently with an ancillary subsystem, which is then manipulated and 
fed back into the system through another coherent interaction (in which case, the information extracted from the system and used in its subsequent manipulation is `quantum information'). See Fig.~\ref{types} for a pictorial representation of this distinction.

\section{\Large \bf \textsf{Feedback control of quantum continuous variables}\label{cv}}

In quantum optics, the light quadratures (the quantum counterparts of electric and magnetic fields of electromagnetic waves) 
may be treated as canonical quantum operators $\hat{x}$ and $\hat{p}$, obeying the commutation relation 
$[\hat{x},\hat{p}]=i$. Notice that this is completely equivalent to the description of a motional degree of freedom of a quantum particle 
in first quantisation (with $\hat{x}$ and $\hat{p}$ representing the position and momentum of the particle). 
All the quantum degrees of freedom satisfying canonical commutation relations 
are collectively referred to as `quantum continuous variables', 
due to the continuous spectra of eigenvalues of the canonical operators 
(a terminology which took hold with the advent of quantum information science, where such systems are contrasted with discrete 
variables on finite dimensional Hilbert spaces, like two-level systems, usually known as `qubits').
Quantum continuous variables, due to their analogue nature, are not typical systems of choice for quantum computing architectures 
(although proposals in such a direction do exist \cite{menicucci06}), but they are very interesting for quantum communication and metrology, and can be exploited, in their optical manifestation, for unconditionally secure quantum key distribution \cite{braunstein05,obrien10}. 

It is in fact a metrological application of light fields, namely the interferometric detection of gravitational waves, 
that first sparkled the quest for the production of nonclassical, squeezed states \cite{braginsky80,walls83}. Squeezed states are states 
where the uncertainty on one of the two canonical quadratures is reduced below the vacuum uncertainty, 
implying an obvious advantage for sensing and precision measurements.\footnote{Let us remind the reader that 
the variances (`uncertainties') of canonical quadratures are constrained by the Heisenberg uncertainty principle: $\Delta \hat{x}^2
\Delta \hat{p}^2 \ge 1/4$ (in its simplest formulation).}
To the best of our knowledge, the earliest demonstration of measurement-based feedback control of a quantum system 
was achieved in a purely optical continuous variable setting, to obtain amplitude-squeezed light \cite{yamamoto86}. 
In this experiment, the weak, quantum non demolition measurement of the photon number was achieved via optical Kerr 
effect, and then fed back into the laser pumping rate to achieve squeezing of its fluctuation. 
Sub-Poissonian statistics of the final light, as well as a remarkable degree of squeezing (around $7$ dB, 
measured from the spectral density of the photo-current fluctuations) were thus obtained. 

Such early all-optical developments, and more broadly the general interest in quantum optics 
and in the generation of squeezed light, 
motivated the establishment of a fully quantum theory of continuously monitored systems and feedback control, 
mainly tied to quantum optical settings and to measurement processes accessible to light fields 
(essentially homodyne and heterodyne detections, and their derivations). 
This eventually led to a theory that can be expressed in terms of quantum Langevin equations in the Heisenberg picture or 
`general-dyne' unravellings and stochastic Schr\"odinger equations 
in the Schr\"odinger picture \cite{wiseman10}. 
The first full-fledged theory of continuously monitored and feedback controlled quantum systems is due to Belavkin \cite{belavkin83,belavkin87,belavkin07}. Other early theoretical treatments for quantum optical fields were 
developed in the eighties \cite{haus86,shapiro87}. 
Feedback control based on homodyne detection 
was studied in \cite{wiseman93}, while a general theory encompassing 
stochastic evolution equations for quantum states and quantum Langevin equations for the fields was presented 
in \cite{wiseman94b}. The most general `dyne-unravelling' 
was characterised in \cite{wiseman01}, while 
optimal control for linear quadratic Gaussian state control problems was 
shown to reduce to semi-definite programming in \cite{wiseman05}.
An alternative approach to continuous feedback using optimal state estimation was introduced
in \cite{doherty99}, while the relationships between quantum and classical feedback control 
are further elucidated in \cite{doherty00}. The theory of feedback control was extended 
from all-optical settings to include atoms in \cite{steck04}. 
Quite interestingly, Ref.\ \cite{ahn02} elaborates on the relationship between quantum error correction schemes 
and feedback control.

The interest in the production of squeezed light has not dwindled since the eighties.  
Quite on the contrary, it has been further strengthened by the clarification,
in the context of continuous variable quantum information, of the 
relationship between squeezing, Einstein-Podolski-Rosen correlations, and quantum entanglement
\cite{wolf03,adesso07}. 
The feedback-assisted 
generation of quantum squeezing by homodyne feedback had already been suggested in 
the early days of feedback control theory \cite{wiseman94a}. The generation of 
steady-state optical continuous variable entanglement in the framework of general-dyne unravellings 
was instead first considered in \cite{mancini06,mancini07}, where an homodyne-based strategy was shown to enhance
the performance of a parametric oscillator 
(see the end of Sec.~\ref{deep} for analogous studies with discrete quantum optical variables). 
Such a scheme was proved to be optimal among all possible general-dyne unravellings in \cite{serafozzi10}. 
More recently, these studies have been extended to incorporate thermal noise \cite{genoni12a}, and to 
consider the generation of entanglement between output travelling modes \cite{nurdin12}
(`out-of-loop' quantum entanglement, as opposed to `in-loop', intra-cavity steady-state results).
The generation of nonclassical superpositions by continuous feedback in optical settings 
has also been investigated \cite{negretti07}.
Notwithstanding the undeniable theoretical and applied appeal of squeezed light,
the practical application of feedback techniques to its generation in all-optical set-ups 
has not been pursued after the aforementioned early attempt by Yamamoto, Imoto and Machida \cite{yamamoto86}. 
This is arguably due to the fact that spontaneous parametric down conversion \cite{wu86,ou92a,ou92b}
has proven sufficient for most purposes so far \cite{braunstein05},
discouraging most experimental groups from dealing with the 
technical difficulties involved in the implementation of measurement-based feedback loops. 
As we shall see, some such set-ups have instead recently started to tread the path of coherent feedback control.

There are, however, two areas, quantum non demolition atomic quantum memories and 
cooling of quantum mechanical oscillators, where the 
measurement-based feedback control of continuous variables 
has found favourable ground for experimental implementation.

\section{\Large \bf \textsf{Continuous variable QND memories}}

A well known approach to the storage and retrieval of the quantum state of a travelling light field 
employs the collective interaction of light with atomic ensembles contained in room temperature 
vapour cells \cite{hammerer10}. 
Each atom in the ensemble couples to light via a well defined optical transition, whose two levels constitute 
an effective two-level system (usually referred to as a `pseudo-spin', in analogy with 
two-level spin $\frac12$ electronic systems). Now, for a macroscopic number of atoms whose two levels 
are highly polarised ({\em e.g.}, when almost all atoms are in the ground state), the collective degree of freedom 
defined as the sum of all the pseudo-spin operators is well approximated 
by a pair of canonical operators,
by virtue of the so-called Holstein-Primakoff transformation \cite{kurucz10}. 
The continuous variable state of light crossing the cell can hence be mapped into the ensemble, 
and subsequently retrieved. This set-up allowed for the demonstration of genuine quantum storage of 
coherent states \cite{julsgaard04}, quantum teleportation between light and matter \cite{sherson06}, and coherent quantum storage 
of squeezed and entangled states \cite{jensen11}.
The development of effective quantum memories is a key step towards the operation of 
quantum repeaters, in turn a stringent requirement on the road towards realistic, long-range 
quantum communication \cite{simon10}, and the optimisation of such devices has hence attracted 
much experimental and theoretical work \cite{choi08,lvovsky09,hulya11}.

Although the coherent interaction would already correlate the atomic ensemble  
with the light passing through it, the operation of these continuous variables memories 
is actually optimised by a measurement-based feedback loop. 
The formal description of this feedback scheme is rather simple and illustrates very well the 
action of homodyne feedback on quantum fields. It is therefore worthwhile to sketch it here
(see \cite{hammerer10} for details).
In the QND (quantum non-demolition) regime, light interacts with the atoms in a 
lambda system configuration, whose excited level is adiabatically eliminated 
resulting in an effective `Faraday' interaction between the atomic ($\hat{X}_A$ and $\hat{P}_A$) 
and light ($\hat{X}_L$ and $\hat{P}_L$) quadratures  
proportional to $\hat{P}_L \hat{P}_A$. Such an interaction is also referred to as QND, 
because it would in principle allow one 
to monitor the atomic quadrature $\hat{P}_A$ by measuring the light quadrature $\hat{X}_L$. 
This opportunity is in a sense exploited in the feedback loop, as we shall explain below. 
After one passage of the light mode through the ensemble, the Heisenberg dynamics relating the 
input light quadratures $\hat{X}_L^{in}$ and $\hat{P}_L^{in}$ and the initial atomic 
quadratures $\hat{X}_A^{in}$ and $\hat{P}_L^{in}$ to the output light quadratures 
$\hat{X}_L^{out}$ and $\hat{P}_L^{out}$ and final atomic quadratures $\hat{X}_A^{out}$ and $\hat{P}_A^{out}$ 
is described by the following equations:
\bea
\hat{X}_{L}^{out} &=& \hat{X}_{L}^{in} + \kappa \hat{P}_A^{in} \; , \nonumber \\
\hat{P}_{L}^{out} &=& \hat{P}_{L}^{in} \; , \nonumber \\
\hat{X}_{A}^{out} &=& \hat{X}_{A}^{in} + \kappa \hat{P}_L^{in} \; , \nonumber \\
\hat{P}_{A}^{out} &=& \hat{P}_{A}^{in} \; , \nonumber
\eea
where $\kappa$ is a coupling strength depending on passage time and other dynamical parameters.
As apparent, only the quadrature $\hat{P}_{L}^{in}$ is somehow mapped by the dynamics to 
an atomic quadrature, while an ideal mapping would require $\hat{X}_{A}^{out} = \hat{P}_{A}^{in}$
and $\hat{P}_{A}^{out} = -\hat{X}_{A}^{in}$ (which is equivalent to perfect mapping up to 
an irrelevant optical phase). Since such a mapping cannot be realised with the input-output relationships above, 
the memory operation is improved by feedback control: The output light quadrature $\hat{X}_{L}^{out}$ is monitored 
(through a measurement equivalent to homodyne detection), and the measurement result is then fed back 
as a linear driving with gain $g$ on the atomic quadrature $\hat{P}_A$. 
It can be rigorously shown that this corresponds to the operatorial 
transformation $\hat{P}_A\rightarrow\hat{P}_A+g\hat{X}_{L}^{out}$. By applying such a transformation on the
input-output relationship for the atomic operators above, and setting $g=-1/\kappa$, one obtains:
\bea
\hat{X}_{A}^{out} &=& \hat{X}_{A}^{in} + \kappa \hat{P}_L^{in} \; , \nonumber \\
\hat{P}_{A}^{out} &=& \hat{P}_{A}^{in} + g\hat{X}_{L}^{in} + g\kappa \hat{P}_A^{in} = 
-\frac1\kappa\hat{X}_{L}^{in} \; . \nonumber
\eea
The mapping is thus effected, up to a coherent squeezing factor $\kappa$, and to additional noise 
deriving from the uncertainty on the initial atomic quadrature $\hat{X}_{A}^{in}$. 
The latter could be in principle reduced (typically by preliminary feedback control on the 
atomic ensemble \cite{{kurucz10},thomsen02,vanderbruggen11}), 
in which case $\kappa\simeq 1$ would achieve near perfect mapping.

Atomic QND memories, operating at room temperature 
according to the principles sketched above, surpassed the quantum threshold
(beating any possible classical `measure and prepare' strategy 
\cite{braunstein00,hammerer05,owari08}) in the storage of both coherent 
and displaced squeezed states \cite{julsgaard04,jensen11}. 
Note that feedback-assisted QND memories are an instance of operator, rather than state, control 
(where the ideal operation is the swap between light and atomic ensemble). 
This is to the best of our knowledge the only implementation of feedback assisted 
quantum operator control to date. 
In the context of quantum memories, it is worth mentioning that 
dedicated theoretical studies, such as \cite{ticozzi06}, 
have been devoted to the suppression of dynamics by feedback control 
(see also \cite{wootton12} for a review on error correcting techniques for quantum memories).

\section{\Large \bf \textsf{Quantum feedback cooling}\label{cooling}}

The past ten years have seen a widespread and intense effort by the physics community to achieve ground state cooling of 
massive harmonic oscillators. Such a line of research, drawing its origins from gravitational wave detectors, holds 
considerable promise for applications in precision sensing \cite{geraci10}, in the realisation 
of quantum memories and transducers (devices coupling different quantum degrees of freedom \cite{stannigel10}),
in the test of fundamental state reduction models \cite{ghirardi86,penrose96}, 
and in the exploration of the quantum to classical boundary \cite{marshall03,kleckner08}. 
The most promising avenue towards the realisation of genuine quantum behaviour with massive 
degrees of freedom is certainly opto-mechanics, a paradigm whereby a material harmonic 
oscillator (typically a suspended mirror) is coupled to cavity light modes, which can be used by various techniques 
to drain its energy out, and thus achieve cooling \cite{kippenberg08,marquardt09}.
Cooling being just a specific form of state control, where stabilisation in the presence of thermal noise is clearly desirable, 
its application in this context has obviously been thoroughly considered in both theory and practice.
In tie with the context of this article, we shall only restrict our discussion to {\em active} feedback control, and will 
disregard what the opto-mechanical literature has at times referred to as 
`passive' feedback cooling \cite{gigan06,arcizet06a}.\footnote{Such schemes, which are proving very successful, 
have been since understood 
to be equivalent to standard cavity cooling, in the sense that they use the cavity to preferentially 
scatter blue detuned photons, and hence extract energy from the harmonic oscillator in the process.}

\begin{figure}[t!]
\includegraphics[scale=.5]{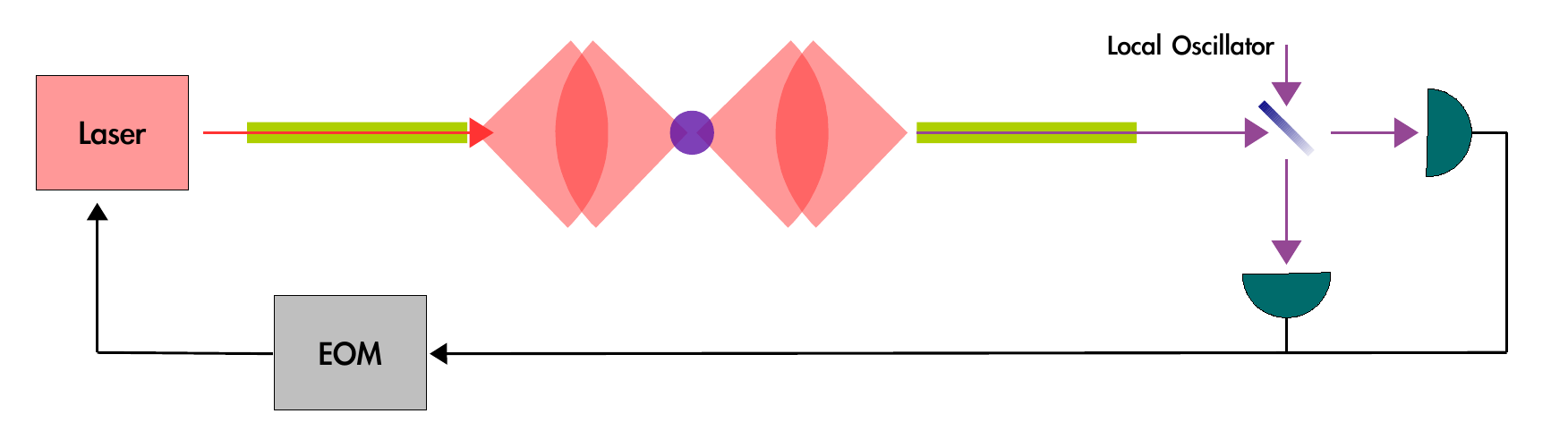}
\caption{\small Simplified schematics of a cold damping feedback loop in opto-mechanics: 
the cavity output is monitored by phase-sensisitve homodyne detection 
at the frequency of a cavity mode which is opto-mechanically coupled to a material oscillator 
(in this particular depiction, a trapped bead in a levitating set-up \cite{chang10,barker10,romeroisart10}). 
The measured current is then 
fed back to the laser through an electro-optical modulator, 
controlling the position and slope of the trapping potential based on 
the homodyne monitoring.
 \label{coldamping}}
\end{figure}

Although similar techniques had long been mastered in the classical regime \cite{ashkin77}, 
the idea of using feedback cooling in opto-mechanics was pioneered in the late nineties \cite{mancini98}, 
and the so called `cold damping' technique (see Fig.\ \ref{coldamping}), where the velocity of the oscillator is continuously tracked by 
phase-sensitive homodyne measurements on the outgoing light and then a negative driving force
is applied on it, simulating friction, was demonstrated soon afterwards \cite{cohadon99}.
An overview of early approaches to feedback cooling in the opto-mechanical context may be found in \cite{vitali02}, 
while a detailed comparison between cavity cooling and cold damping, including optimal operating regimes, 
may be found in \cite{genes08}. 
There, it is shown that cold damping is favourable in the bad cavity limit. In fact, one of the most convincing applications of 
cold damping to date was achieved in the cooling of a motional degree of freedom of a single trapped ion, 
where average occupations down to $3$ excitations seem to be within reach \cite{bushev06}. 
Such a set-up enjoys the major advantage, 
over standard opto-mechanics, of not involving a cavity for light. Phase sensitive detection, analogous to homodyning, was 
implemented by modulating the ion's fluorescence with a retro-reflecting mirror. Active control was actuated by acting on the trap electrodes, in general a notable control possibility available in ion traps \cite{heinzen90,serafozzi09a,serafozzi09b,brown11}.
In cavity mediated set-ups, 
cold damping of a micro-mechanical resonator down to $5$ K was demonstrated in \cite{arcizet06b}, and 
improved to $100$ mK in \cite{kleckner06}. Soon afterwards, an oscillator of about $1$ g was cooled down to the 
mK region by cold damping, as reported in \cite{corbitt07}.

Analogous techniques, employing active measurement-based feedback control and often called `parametric cooling' 
(where the control parameter is generally different from case to case), were also recently applied to cool 
cantilevers \cite{poggio07}, 
gravitational wave detectors \cite{vinante08}, electromechanical quartz resonators \cite{jahng11} 
and levitated micro-and nano-particles \cite{li11,gieseler12}, 
although none of such systems have yet been brought into the deep quantum regime by cold damping so far.
Feedback cooling and control schemes have also been proposed and studied in detail for Bose-Einstein condensates 
\cite{wilson07,szigeti09,szigeti10}.

\section{\Large \bf \textsf{State control by measurement-based feedback in the deep quantum regime}\label{deep}}

With few exceptions, the experimental achievements covered so far, while genuinely `quantum' in that they are subject 
to measurement back-action and act on quantum variables, operate in a semi-classical regime, where the quantum fluctuations 
are small with respect to the average fields. We will now move on to demonstrations in the deep quantum regime, where 
fields and matter are controlled down to a few quanta. 

Cavity QED systems, where high finesse cavity light, typically in the microwave region, 
is strongly coupled with flying or trapped atoms,
is arguably one of the most controllable quantum systems, and has hence an outstanding pedigree in the demonstration 
of quantum information 
and control primitives \cite{raimond01,miller05,walther06}.\footnote{Arguably, the advantages 
of standard QED set-ups are offset
by difficulties in terms of scalability, especially with respect to integrated optical components of recent development in the 
linear optical arena \cite{politi09}.} 

It should therefore come as no surprise that the first demonstration of deep quantum feedback control techniques  
was achieved with a cavity QED set-up \cite{smith02} (see also \cite{reiner04} for an extended study). 
In this early experiment, a thermal beam of $^{85}{\rm Rb}$ atoms was sent through a cavity weakly driven at a 
frequency resonant with an atomic transition, thus coupling the atoms strongly with the cavity field but keeping the 
maximum number of excitations inside the system very low (around two). The state was then conditioned upon the 
measurement of a photon leaking out of the cavity mirrors, and then actively controlled by applying a change in the driving intensity 
after a pre-determined time $T$ from the measurement (estimated thanks to the detailed knowledge of the atom-light dynamics). 
It was hence possible to demonstrate, in terms of the measured intensity correlation function $g^{(2)}$, the freezing of the 
conditional dynamics for the duration of the control pulse, and then to subsequently resume the Rabi oscillations of the system. 
Although only triggered by a single measurement outcome (detection of a photon), and hence intrinsically non-deterministic, and 
involving a single pre-determined operation (such that the only factor determined by the measurements in real time was the timing of the intensity change), this experiment was the first to single out and partially control a conditional state (a `quantum trajectory') 
in the full quantum regime. Notice that the measurement is effectively weak, in that it is based on the detection of an output photon which had been previously coupled to the internal field through the mirrors. 
However, this experimental demonstration did not go so far as to stabilise a quantum state through feedback control.

Further advances in feedback state control have made use of qubits (two-level quantum systems) encoded in the polarisation of travelling photons. These linear optical set-ups have been extremely successful since the early days of quantum information, enjoying low decoherence rates and ample possibilities for coherent manipulations on single qubits. They were in fact employed in the earliest demonstrations of quantum teleportation \cite{bouwmeester97,boschi98}. 
More recently, the implementation of entangling quantum gates on two such qubits has also 
been reported \cite{langford05,kiesel05,okamoto05}, including controlled-Z (CZ) gates, which will be relevant in what follows.     
In \cite{gillett10}, the approximate recovery of a polarisation state after the application of a noisy channel 
has been demonstrated by a single shot quantum feedback procedure. The qubit was prepared in one of two possible
non-orthogonal states, and underwent dephasing through a probabilistic Z gate. It was then observed by 
the application of an entangling CZ with another meter qubit, which was then projectively measured, thus effecting the weak measurement on the system qubit. A tuneable coherent coefficient in the initial state of the meter qubit 
allowed one to range from (strong) non-demolition projective measurements on the system qubit 
to the trivial no measurement case, spanning a whole class of dichotomic (two outcomes) POVMs in between. 
After the dichotomic measurement, the feedback loop was closed by applying a unitary operation 
whose sign was dependent on the measurement outcome (a rotation in the Y axis of the Bloch sphere, 
conditionally implemented by means of a Pockels cell). By tomographic reconstruction of the final states, this  
experiment showed that, in this particular setting, weak measurements outperform ``do nothing'' 
and projective measurements-based strategies in terms of average overlap between 
the reconstructed and initial quantum states. 
While not yet capable to stabilise a target state, and only applying to the single shot recovery of 
specific states subject to specific forms of discrete noisy channels, this demonstration constituted a remarkable 
advance in terms of the control demonstrated in the implementation of weak measurement processes.
It should be noted that a similar scheme with qubits in a linear optical set-up, 
based on analogous measurements by ancillary projection and feed forward, had been previously realised 
to obtain minimum disturbance weak measurements \cite{sciarrino06}, 
whose realisation had already been envisaged in \cite{genoni05}.\footnote{It is also worth remarking that 
similar examples of single-shot `feedforward' manipulations are also employed 
in all forms of quantum teleportation, both with polarised photons 
\cite{bouwmeester97,boschi98,giacomini02} and 
in the continuous variable regime \cite{furusawa98}. 
This class of experiments, centred on the single-shot manipulation of quantum systems 
for the transmission of quantum information, is rather far in spirit from real-time feedback control, and has hence 
been deemed outside the scope of the present discussion.} 

For the first experimental demonstration of real time steady state control by measurement-based feedback,
we have to turn back to a cavity QED set-up, one with a remarkable track record 
of experimental achievements over the last 20 years \cite{sayrin11,raimond01,gleyses07,guerlin07,deleglise08}. 
The experiment reported in \cite{sayrin11} consists in the stabilisation, by iterated feedback loops, 
of a photon number (Fock) state in a superconducting microwave cavity. 
The meter subsystem, in this case, is represented by a beam of rubidium atoms crossing the cavity 
and interacting dispersively with the cavity light field, such that the global atom-cavity light state 
acquires phases depending on the photon number in the cavity. A subsequent phase sensitive 
detection of the atom then reveals partial information about the photon number in the cavity, and such information 
is used to steer the cavity field state to the desired number state by coupling the cavity to a classical 
current whose intensity and phase can be modulated to optimise the overlap with the desired target state.
Although the implementation of this experiment is clearly very delicate given the high degree of coherent control demanded, 
and requires to take several noise factors and potential imperfections into account, the idealised underlying 
theoretical treatment is rather simple, and worth describing in some detail 
(see \cite{amini12} for a rigorous discussion).
The unitary operation acting on the atom-light system 
during each atomic passage through the cavity is given by $U=\exp(-i\varphi_0(\hat{N}+1/2)\sigma_z)$, 
where $\varphi_0$ is phase determined by passage time and coupling strength, $\hat{N}$ is the cavity field 
number operator (at the near resonant frequency), while $\sigma_z$ acts on the ground ($\ket{g}$) 
and excited ($\ket{e}$) atomic states according to $\sigma_z\ket{e}=\ket{e}$, $\sigma_z\ket{g}-\ket{g}$. 
The cavity is sandwiched by an atomic Ramsey interferometer where the atomic states are coherently rotated before 
and after crossing the cavity, with a tuneable relative phase $\varphi_r$: assuming that the initial state of each atom 
is $\ket{g}$, the first auxiliary cavity turns the atomic state into $(\ket{g}+\ket{e})/\sqrt{2}$. 
The initial global cavity light-atomic state is then given by 
$$\ket{\psi_0}=\frac{1}{\sqrt{2}}\sum_{n}\psi_n\ket{n}\otimes\left(\ket{g}+\ket{e}\right) \; ,$$ 
where the initial state of the cavity is assumed to be any pure state (it is actually a coherent state in the experiment, 
but such a detail is irrelevant to our current argument), and $\ket{n}$ stands for an eigenvector of the number operator: 
$\hat{N}\ket{n}=n\ket{n}$. The interaction during passage through the cavity entangles the state of the cavity field with that of light, giving the new global state: 
$$U\ket{\psi_0} = \frac{1}{\sqrt{2}}\sum_n \psi_n \left({\rm e}^{i\varphi_0 (n+1/2)} \ket{n,g}+{\rm e}^{-i\varphi_0 (n+1/2)} 
\ket{n,e}\right)$$ 
(with $\ket{n,j}=\ket{n}\otimes\ket{j}$ for $j=g,e$). In the second Ramsey cavity the state of the atoms is rotated again, 
according to $\ket{g}\rightarrow(\ket{g}+\ket{e})/\sqrt{2}$ and $\ket{e}\rightarrow(-\ket{g}+\ket{e})/\sqrt{2}$ 
(setting for simplicity $\varphi_r=0$), so that the global state turns into 
$$\sum_n \psi_n \left(i\sin(\varphi_0(n+1/2)) \ket{n,g}+\cos(\varphi_0(n+1/2)\right) \ket{n,e}) \; .$$ 
It is now apparent that a measurement of the atom in the $\{\ket{g},\ket{e}\}$ basis 
(realised by a field-ionisation detector after the Ramsey interferometer) would 
makes the atomic state collapse into the (unnormalised) state 
$$\sum_n \psi_n \sin\left(\varphi_0(n+1/2)\right) \ket{n} \quad {\rm (for\; outcome} \; g {\rm )} \, , $$ 
or 
$$\sum_n \psi_n \cos\left(\varphi_0(n+1/2)\right) \ket{n} \quad {\rm (for\; outcome} \; e {\rm )} \, . $$ 
This can be expressed by stating that the projective measurement of the atomic meter effects the dichotomic POVM described by the operators $M_g=\sin(\varphi_0(\hat{N}+1/2)+\varphi_r)$ 
and $M_e=\cos(\varphi_0(\hat{N}+1/2)+\varphi_r)$ (where the tuneable interferometric phase $\varphi_r$, 
omitted for simplicity in the derivation above, has been re-instated). 
After each measurement, given the outcome $j=e,g$, the quantum state in the cavity 
is updated as per $\varrho \rightarrow M_j \varrho M_j^\dag/{\rm Tr}(\varrho M_j^\dag M_j)$, and 
a CPU-based controller estimates the value and sign of the optimal optical displacement to apply to the 
cavity field in order to maximise the figure of merit $\bra{n_T}\varrho\ket{n_T}$ at each step. 
Here, $\ket{n_T}$ is the target number state.

This experiment demonstrated final fidelities of about $0.8$ for Fock states $\ket{1}$, $\ket{2}$, $\ket{3}$ 
and $\ket{4}$, reached in times of the order of $50$ ms 
in the presence of decoherence due to photon losses and thermal photons 
(even in a cryogenic environment, the average number of thermal photons 
at the frequencies adopted is still $0.05$, and hence not completely negligible). 
The accuracy of the experiment is still limited, first and foremost, by the limited efficiency of the 
atomic detectors (around $35$\%), as well as by the irregularities in the atomic beam 
(atoms might be missing every now and again). Even so, this demonstration constitutes a notable advancement on the application of feedback control to a quantum optical set-up, and well exemplifies the 
potentialities of such control techniques. It is probably relevant to emphasize that  
Fock states, being deeply quantum (as opposed, for instance, to intense coherent states, whose statistics 
can be mimicked to a good extent by classical light), are a very desirable ingredient in the implementation of 
several quantum communication protocols, for instance whenever 
a reliable single photon source is required \cite{vanenk98,lo99}. 
Fock states are also extremely fragile in the face of decoherence \cite{guerlin07,wang08,brune08}
(see \cite{serafozzi04} for a theoretical treatment of the decoherence of number states and superpositions thereof), 
and the quest for their generation on demand has been long 
and riddled with difficulties \cite{varcoe00,hofheinz08,hofheinz09,eisaman11}. 
A feasible scheme which allow for their stabilisation in the presence of quantum noise, while still imperfect, is 
therefore a significant step forward in quantum state preparation and control. Such measurement-based schemes 
for the generation of nonclassical schemes are to be contrasted with measurement-free ones, such as those suggested in \cite{lofranco07,lofranco10}.

The feedback scheme of \cite{sayrin11} has been further refined, utilising the same set-up, in a very recent experiment \cite{zhou12}.
Here, the weak detection is still implemented through flying Rydberg atoms interacting with the cavity field, but 
the actuator is capable of changing, in real time and for variable time intervals, 
the form of the atom-light interaction inside the cavity, switching from 
dispersive to resonant couplings, and thus correcting directly for the leakage of one quantum inside the cavity due to loss 
(the atoms are also previously pumped in the excited state while the interaction is resonant, 
such that they can yield an excitation to the field). This allowed for demonstrating 
the stabilisation of number states up to $n=7$ and 
is, arguably, the most explicit 
demonstration of measurement-based quantum feedback state control to date (in that the coherent Hamiltonian control is not limited to applying additional operations, but act directly on the form of the meter-system coupling in order to counteract decoherence).

On the theoretical side, it is worth noting that schemes to achieve and preserve highly entangled steady-states between 
internal atomic degrees of freedom, analogous to those discussed for continuous variables in Sec.~\ref{cv}, 
have been proposed, analysed and extended to multipartite set-ups in \cite{carvalho07,carvalho08,stevenson11}.

\section{\Large \bf \textsf{Adaptive measurements}}

Alongside their potential for the control of quantum states and operations, feedback techniques have also found application 
in precision measurements and quantum metrology in optical set-ups. 
It has long been known that, because of the uncertainty 
relations, quantum mechanics imposes bounds on the attainable precision of quantum measurements \cite{helstrom76}. 
Assuming the vacuum (or ground state, for a mechanical harmonic oscillator) uncertainty as the fundamental 
unit of error, one obtains the so called `standard quantum limit' to the statistical precision of a measurement
process, characterised by a $1/\sqrt{N}$ scaling of the precision 
($N$ being the number of ``resources'', such as repetitions of the measurement). 
It has however also been noted that such a limit can be beaten by adopting squeezing or, more subtly, quantum entanglement, a remark that has given rise to the subfield of quantum metrology \cite{giovannetti04} 
(see also \cite{giovannetti11} for a more recent review). 
Note that, although the standard quantum limit can be beaten, 
the ultimate precision allowed by quantum mechanics is still bound by the so-called Heisenberg limit, 
which sets the ultimate scaling of $1/N$ on the precision compatible with the uncertainty principle. 

Besides the recourse to optimised nonclassical states, 
it has also long been known that the adoption of feedback techniques  
enhances the precision of quantum measurements. The notion of `adaptive' quantum measurement was first formulated 
in a quantum optical setting by considering the estimate of the phase of an optical mode \cite{wiseman95},
where it was pointed out that a homodyne scheme where the latest estimate of the phase
is fed back in real time to continuously adjust the local oscillator phase would beat 
a non-adaptive heterodyne strategy (the method of choice for phase estimation until then). 
The superiority of the adaptive homodyne scheme for optical phase estimation 
was experimentally demonstrated with weak coherent states in \cite{armen02}.
A number of theoretical and experimental developments followed.
An adaptive technique for the estimate of an 
interferometric phase difference was suggested in \cite{berry00}. 
The advantage offered in the estimate of noisy, fluctuating optical phases by adaptive strategies
for coherent and (broad- and narrow-band) 
squeezed input states was further analysed in \cite{berry02} and \cite{berry06}.
The standard quantum limit in phase estimation was eventually beaten 
by using non-entangled single photon states and an algorithmic adaptive technique, 
which allowed essentially to reach the Heisenberg limit, as reported in \cite{higgins07}.
Continuous adaptive measurements were also applied to phase estimation by quantum smoothing, a 
`time-symmetric' post-processing technique where the estimate at time $t$ is inferred based on both past
(up until time $t$) and future (after time $t$) measurement records: such a technique allowed for a reduction by a factor $2.24$ with respect to the standard quantum limit achievable by causal filtering \cite{wheatley10}.
An adaptive interferometric scheme along the lines of \cite{berry00} has been recently applied \cite{xiang11} to achieve phase estimation below the shot noise limit -- yet another constraint implying a $1/\sqrt{N}$ scaling precision in photon counting, due to the discrete nature of photons -- by utilising heralded entangled states with a given number of photons
(a generalisation of the NOON states, known to be optimal for phase estimation at the Heisenberg limit 
\cite{lee02,giovannetti04}). 
It must be noted that such a scheme required some degree of post-selection upon the final detection.

With the notable exception of \cite{xiang11}, which however made use of post-selection, 
most of the results discussed so far refer to phase sensing, 
where the phase to be estimated is known to lie within some given interval. 
The first demonstration of sub-Heisenberg phase tracking 
({\em i.e.}, the estimate of a completely unknown noisy optical phase), 
with no sort of data post-processing nor theoretical compensation for imperfections, 
resorted to squeezed states and adaptive homodyne measurements 
where the phase of the local oscillator was updated in real-time by Kalman filtering, 
which also provided the optimal estimate of the phase \cite{yonezawa12}.\footnote{A Kalman filter \cite{simon06} 
is a standard recursive algorithm for classical and quantum state estimation, 
where the updated estimate at any time only depends on the estimate at the previous time-step and on the latest measurement result.} 
The error on the estimate of the randomly fluctuating phase was around $15$\% 
below what could be obtained with coherent input states, an impressive
achievement assuming no a priori information on the noise. 

Adaptive quantum state estimation, where the objective is the reconstruction of a quantum state given 
a certain number of available copies, and
whose efficiency was theoretically proven in \cite{fujiwara06}, 
has also been very recently demonstrated in the laboratory with a linear optical set-up \cite{okamoto12}.

Somewhat related to these developments are the demonstrations of state discrimination
by adaptive measurements.\footnote{By `state discrimination', 
one refers to the task of distinguishing between different non-orthogonal quantum states 
with limited available resources ({\em e.g.}, number of copies of the state). 
Note that no single-shot measurement can deterministically distinguish between 
two non-orthogonal quantum states.} 
Real-time feedback enhanced discrimination between optical coherent states was experimentally demonstrated in \cite{cook07}
while, in \cite{higgins09}, adaptive measurements 
have been shown to beat locally optimal 
({\em i.e.}, optimised over single-shot measurements, see \cite{acin05} for a theoretical discussion) 
strategies with single photons undergoing depolarising noise. 
These experiments are strictly related to the results presented in \cite{gillett10}, where however a final 
outcome-dependent manipulation of the photons were performed, and which have hence been already discussed
as cases of quantum state control (also, it must be emphasized that 
the experiments reported in \cite{gillett10} and \cite{sciarrino06}, while making use 
of a closed-loop for the realisation of each individual run, are not based on properly `adaptive' detections). 
Broadly speaking, the key advantage offered by feedback control in the context of quantum estimation and metrology is 
the possibility of gaining a quantum advantage with limited recourse to fragile and elusive resources (such as NOON states).
Finally, let us also draw the reader attention on an interesting systematic approach to adaptive measurements 
based on machine learning, introduced in \cite{hentschel10}.

\section{\Large \bf \textsf{Coherent feedback control}}

Measurement-based quantum feedback is based on the extraction of {\em classical} information from a quantum system to steer its subsequent control. Even when the latter is implemented through coherent manipulation, this approach is in principle sub-optimal, because the {\em quantum} information contained in the system is for the most part not used to control 
it.\footnote{Let us remark that there exists no procedure capable of converting a single copy of an
arbitrary quantum state into classical information.} This observation gave rise to the notion of coherent feedback control of a quantum system \cite{lloyd00},
which had already been envisaged in the nineties within a quantum optical scenario, under the name of `all-optical' 
feedback \cite{wiseman94c}. 
Coherent control extracts part of the quantum information contained in the system by coherent interaction with an ancillary
system; such a quantum information is subsequently coherently manipulated and then fed back to the system through coherent interaction. 
Notice that this approach is obviously free from measurement back-action noise on the system.
Although many experiments have adopted 
procedures that can be partially assimilated to coherent feedback control, the first explicit demonstration of quantum state control based on a coherent feedback loop was given in an NMR set-up \cite{nelson00}.

In quantum optics, coherent feedback control can be typically implemented by letting the output of a cavity 
undergo some controlled coherent evolution and then feeding it back as a cavity input. 
These situations can be described by the so-called input-output formalism \cite{collett84}.
A comprehensive theoretical treatment of very general multimode systems encompassing 
all linear canonical field transformations (also known as `Bogoliubov', or `symplectic' transformations, 
including beam splitting, squeezing and phase shifting) and coherent feedback loops 
with finite delays based on input-output dynamics -- as well as series (or cascade) connections -- has been recently 
developed within the context of quantum quantum feedback networks \cite{gough08,gough09a,gough10},
partially building on an existing approach \cite{yanagisawa03a,yanagisawa03b}.
The optimisation of the production of steady state squeezed light by coherent feedback was 
investigated in \cite{yanagisawa03b} and \cite{gough09b}, and experimentally demonstrated in \cite{iida12},
where the squeezing of an OPO (optical parametric oscillator) was enhanced from $-1.64$ dB to $-2.20$ dB 
by means of a coherent feedback loop where input coherent light was mixed with the OPO cavity output 
at a controlling beam splitter and then fed back into the OPO, in order to amplify the parametric 
oscillation. 
By considering a similar scheme, 
the enhancement of continuous variable optical entanglement by coherent feedback was quantitatively 
studied in \cite{yan11}. 
Very recently, it has also been shown that coherent feedback would improve the 
feedback cooling of mechanical oscillators \cite{hamerly12a,hamerly12b}, a perspective which may open up new pathways 
for opto-mechanics (see Sec.\ \ref{cooling}).

Let us also mention that, over the past few years, considerable theoretical work has been devoted to the understanding of the 
open- and closed-loop coherent controllability of infinite dimensional bosonic systems, 
by $H^{\infty}$ \cite{james08,maalouf11}, linear \cite{nurdin09} and symplectic \cite{genoni12b} controllers. 
Such theoretical results are of direct interest to the coherent control of quantum optical set-ups.

\section{\Large \bf \textsf{Outlook}}

The coherent control and manipulation of quantum systems is growing more and more central to
quantum optics and, more broadly, to the whole physics community \cite{nobel12}.
In light of the promise feedback methods hold, the refinement, and extension to other fields and tasks, of the 
schemes designed and demonstrated so far is hence envisageable in the short and medium term.
As a relevant example, feedback techniques based on strong measurements seem to be ripe for application 
with transmon qubits coupled to superconducting resonators (mimicking a QED system) \cite{mirrahimi12}.
Outstanding opportunities are also still to be explored with coherent control strategies, 
such as their application to nano-photonics \cite{kerckhoff10,mabuchi11}.

\section*{\large \bf \textsf{Acknowledgments}}
A number of colleagues have helped me, directly or indirectly (by collaboration on topics relevant to this article), 
during the completion of this manuscript, 
among whom I would like to mention M.\ Barbieri, D.\ Vitali, F.\ Sciarrino, A.R.R.\ Carvalho, S.\ Mancini and P.\ Barker. 
Special thanks go to M.G.\ Genoni, whose assistance at all stages of this endeavour was invaluable.


\end{document}